\documentclass[english,reprint,onecolumn,amsmath,amssymb,notitlepage]{revtex4-1}

\newcommand{\sio}{SiO$_2$ }
\newcommand{\vg}{$V_g$ }
\newcommand{\vd}{$V_d$ }

\usepackage{comment}
\usepackage{graphicx}
\usepackage{color}
\usepackage{units}
\usepackage[normalem]{ulem}
\usepackage{docs}%
\usepackage{bm}%
\usepackage[colorlinks=true,linkcolor=blue,citecolor=red]{hyperref}%

\begin{document}

\title{Role of different scattering mechanisms on the temperature dependence of transport in graphene}
\author{Suman Sarkar}
\author{Kazi Rafsanjani Amin}
\author{Ranjan Modak}
\author{Amandeep Singh}
\author{Subroto Mukerjee}
\author{Aveek Bid}
\email{aveek.bid@physics.iisc.ernet.in}
\homepage[\\visit at:]{http://www.physics.iisc.ernet.in/~aveek_bid/aveek.html}
\affiliation{Department of Physics, Indian Institute of Science, Bangalore 560012, India}


%
%

\begin{abstract}
Detailed experimental and theoretical studies of the temperature dependence of the effect of different scattering mechanisms on electrical transport properties of graphene devices are presented. We find that for high mobility devices the transport properties are mainly governed by completely screened short range impurity scattering. On the other hand, for the low mobility devices transport properties are determined by both types of scattering potentials - long range due to ionized impurities and short range due to  completely screened charged impurities. The results could be explained in the framework of Boltzmann transport equations involving the two independent scattering mechanisms. 
\end{abstract}

\maketitle

\section*{Introduction}
In two-dimensional conductors the ratio of transport scattering time $\tau_t$ and the single particle scattering time $\tau_s$ depend very strongly on the nature of the scatterers, the screening properties of the substrate and the relative location of the scatterers with respect to the conducting layer~\cite{PhysRevB.32.8126, PhysRevB.38.10798, PhysRevB.32.8442}.  In conventional semiconductors with a parabolic dispersion the transport scattering time $\tau_t$, in contrast to the single particle  scattering time (or quantum scattering time) $\tau_s$, is largely insensitive to small angle scattering.  It has been shown in high mobility modulation doped GaAs/AlGaAs two dimensional electron gas (2DEG) systems that the spatial separation of the remote ionized impurities and the 2DEG, while ensuring that the large angle scattering of the charge carriers is small, does not rule out small angle scatterings. This results in significantly large $\tau_t$ (and consequently large charge carrier mobilities) but in no way guarantees a large single particle scattering time $\tau_s$. Thus even a relatively high mobility device may suffer from significant broadening of the single particle momentum eigen-levels (resulting from a large Dingle temperature~\cite{PhysRevB.32.8126}~\cite{PhysRevB.37.4299}). In the case of graphene, an additional feature arising due to the chiral nature of the transport is the absence of back-scattering leading to interesting dependence of the scattering rates on the nature and location of impurities~\cite{dassarma.2008}. Specifically, this makes the ratio of $\tau_t/\tau_s$ extremely sensitive to the exact details of impurity type and configuration.  It has been shown experimentally that in graphene the integer quantum Hall states break down at much lower dissipation levels than in GaAs/AlGaAs based devices. It was proposed that this is due to the large quasielastic inter-Landau level scatterings that are aided by the local electric field created by long range charged scatterers ~\cite{PhysRevB.85.165420}. It is clear that the dependence of $\tau_t$ and $\tau_s$ on long range charged impurity scattering and delta potential scattering has important bearings on transport properties of graphene - a topic that has not received adequate experimental attention.
 
Electronic transport properties of graphene, especially the temperature dependence of its resistance, has been studied over a vast temperature range from room temperature down to few tens of milli Kelvin. The emphasis has been on making the cleanest possible devices~\cite{bolotin2008temperature} to extract the intrinsic physics of graphene. In the case of single layer graphene (SLG) devices that were suspended, it has been shown that the linear variation of the resistivity $\rho$ with temperature can be explained by scattering due to acoustic phonon and an unexpected non-linear density dependence of $d\rho/dT$. It is presumed that this is because of screened coulomb scattering at low carrier densities. The picture is much more complicated in the case of the technologically relevant device architecture of SLG devices on substrates.  A peculiar consequence of the linear dispersion relation of SLG is that the resistivity due to acoustic phonon scattering was predicted to be independent of the carrier number density~\cite{PhysRevB.22.904,PhysRevB.76.205423}. This was found to be experimentally valid~\cite{nature2008} in the temperature range below 200K for ultra-clean SLG devices on \sio substrates. Over this temperature range $d\rho/dT$ was independent of number density (gate voltage) and $\rho$ varied linearly with temperature.  At higher temperatures the temperature dependence of resistivity became nonlinear with the nonlinearity increasing sharply as the chemical potential approached the Dirac point. It was proposed that at temperatures below 200~K the dominant scattering was due to acoustic phonons, beyond 200~K remote interfacial phonons (RIP) arising from the \sio substrate also become relevant. Alternately, it was proposed that long range Coulomb scattering alone can explain most of the available experimental data on temperature dependent transport
~\cite{hwangprl.2007,macdonald.2006,Adam20112007}.
This, however required much higher impurity number densities than is experimentally observed. Thus, despite intensive research over the last decade the relevant scattering mechanisms which determines the transport properties have not been unambiguously identified. 

\section*{Results}

\begin{figure}[!h]
\begin{center}
\includegraphics[width=0.75\textwidth]{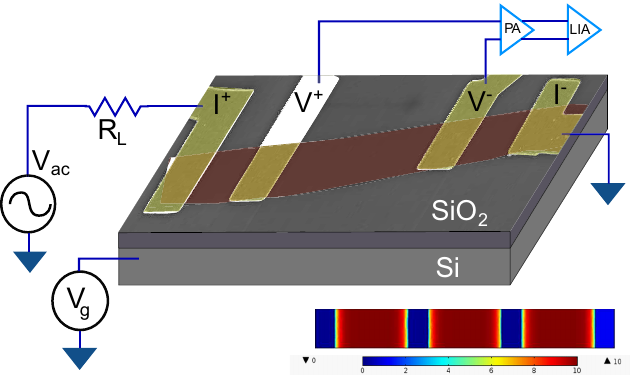} 
{\caption {Schematic of the SLG graphene device based on false colour SEM image of the device g28m6. The SLG was deposited on a 300 nm SiO$_2$ substrate. Here $R_L$ is the ballast series resistance, $V_g$ is the back gate voltage, and $V_{ac}$ is the source-drain bias, PR is the low-noise room temperature preamplifier (SR552) and LIA is the dual channel lock-in amplifier (SR830). The lower part of the image shows the distribution of the potential profile in the conducting channel due to $V_g$  (see text for details). } \label{Fig:figure1}}
\end{center}
\end{figure}

We have studied in detail the temperature dependence of the resistivity of SLG devices on \sio substrate to understand the temperature and gate voltage ($V_g$) dependencies of the scattering mechanism. SLG was exfoliated on Si/\sio substrate and the number of layers was confirmed using Raman spectroscopy. Electrical contacts were made on selected SLG using standard electron beam lithography techniques. Resistivity measurements were carried out on each device at low frequencies using standard 4-probe lock-in techniques. A schematic of the device is shown in figure~\ref{Fig:figure1}.  The measurements were all performed at low frequencies, about 228Hz. The capacitive effect in all cases was negligible as seen from the near zero value of the quadrature component of the voltage across the sample measured simultaneously by the dual channel lock-in amplifier. The current used in these measurements was 100~nA. From the measured thermal conductivity of our graphene devices ($\sim 400$ Wm$^{-1}$K$^{-1}$) we estimate the maximum temperature increase of the SLG device due to Joule heating to be about 10~mK. We have carried out the measurements in multiple devices with different amounts of intrinsic disorder to quantify the effect of various types of disorder on the temperature and $V_g$ dependencies of the resistance.  In this report we present the data from a few representative devices - the details of the device parameters measured at 295~K are given in table~\ref{tab:1}. The devices were subjected to different degrees of cleaning during the lithography process - this resulted in devices whose  mobilities varied by almost than two orders of magnitude. We have quantified the amount of disorder in the different devices through the Ioffe-Regel parameter $k_Fl$, where $k_F$ is the Fermi wave-vector of the SLG and $l$ is the carrier mean free path  - the results are shown in table~\ref{tab:1}. It can be seen that for the device with maximum disorder (g10m6) the value of $k_Fl$ is very close to the Mott-Iofffe-Regel limit for metallic conduction~\cite{iofferegel,PhysRevB.24.7404}.

\begin{table}[!h]
\centering
\begin{tabular}{| r | c | c | c | c | c | c | l | p{2cm}|}
\hline
 device name& $\mu$    & $n_0$  & $length$ & $width$ &$Dirac point (V_d)$ &$dW_1/dT$ &$dW_2/dT$ &$k_Fl$   \\ \hline
   	       & ($m^2V^{-1}s^{-1}$) & ($m^{-2}$) & ($\mu$m)  & ($\mu$m) & ($V$)  & ($K^{-1}$) & ($K^{-1} $) & $ $    \\ \hline
g28m6 	   &2.36     &	0.9e15  &	12.5 &	3.9   &	-8.0 & +ve & 0 & 178 \\ \hline
g30m4      &1.89     &	2.2e15  &	5.0 &	1.4 &	-5.0 & +ve & -ve & 122\\ \hline
g7m5       &1.64     &	3.7e15   &	4.0  &	2.6 &	-8.0 & +ve & -ve & 62 \\ \hline
g37m2      &1.09   &	2.5e15   &	2.4 &	2.4 &	4.5 & +ve & -ve & 43\\ \hline
c2s2       &0.25     &	6.3e15   &	1.8 &	3.5 &	49.0 & +ve & -ve & 14 \\ \hline
g10m6      &0.03    &	1.5e16   &	4.0  &	2.0 &	16.0 & -ve &  +ve & 3 \\ \hline
\end{tabular}
{\caption{Device parameters for a few representative devices discussed in this manuscript. $\mu$ is the charge carrier mobility, $n_0$ is the impurity number density, $dW_1/dT$ and $dW_2/dT$ are the temperature coefficients of the long range Coulomb scattering and short range delta scattering weight factors respectively,  $k_F$ is the Fermi wave-vector, $l$ is the charge carrier mean free path and $k_Fl$ is the Ioffe-Regel parameter. 
\label{tab:1}}}
\end{table}

\begin{figure}
\begin{center}
\includegraphics[width=0.65\textwidth]{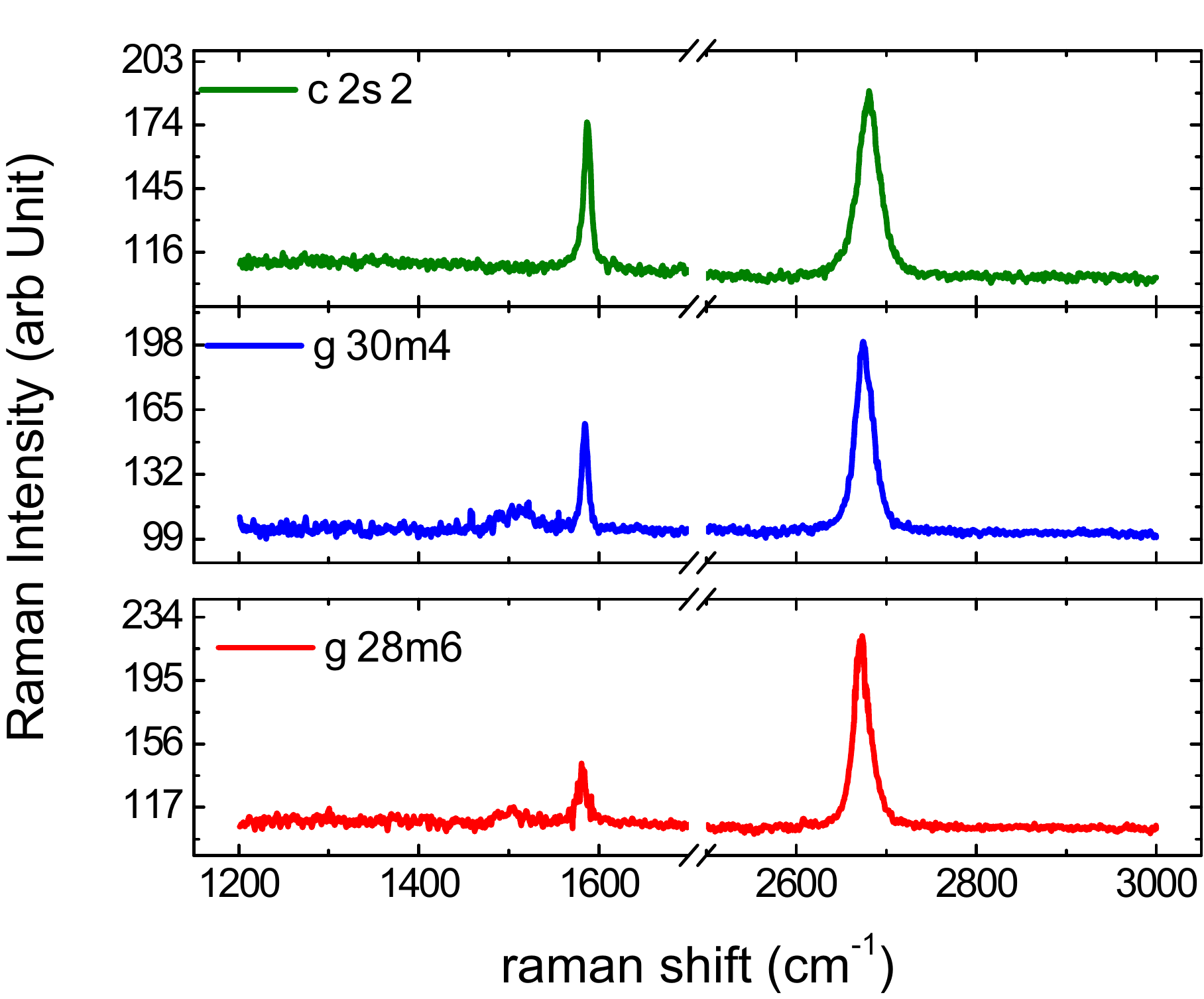}
{\caption{Raman spectrum of the SLG devices after the lithography processes. From the ratio of the intensities of the 2D-peak and the G-peak we can conclude that the device g28m6 is the cleanest while c2s2 is the most disordered. 
\label{fig:figure2}}}
\end{center}
\end{figure}

Figure~\ref{fig:figure2} shows the Raman spectrum for a few of the devices. From the ratio of the intensities of the 2D-peak and the G-peak we can conclude that the device g28m6 is the cleanest while c2s2 is the most disordered. The mobility $\mu$ and number density of charge impurities $n_0$ in these devices were estimated by fitting the $R$-\vg curves measured at each temperature with the relation~\cite{PhysRevB.80.235402,kim2009realization,venugopal2011effective,hsu2011impact}

\begin{eqnarray}
R =  R_c +\frac{l}{w}\frac{1}{e\mu\sqrt{n_0^2+e^2C^2(V_g-V_d)^2}}
\label{Eqn:dassharme}
\end{eqnarray}
\noindent
where $R_c$ is the contact resistance, $l$ the length of the SLG between measurement probes, $w$ is the width of the SLG, $\mu$ is the mobility of the charge carriers, $C$ is the capacitance per unit area of the SLG FET device and $n_0$ is the charge impurity concentration in the SLG device. It should be noted that $R_c$ is not a contact resistance in the conventional sense – it is the resistance of the part of graphene monolayer very close to the contact leads where the effect of the back gate is screened off by the metallic leads. The device thus effectively acts as two resistors in series – a component far enough from the metallic contacts whose Fermi energy (and consequently resistance $R(V_g)$) can be modulated by the gate voltage $V_g$, and another component very close to the contacts whose resistance $R_c$ is essentially independent of the back gate voltage. We have performed extensive three-dimensional electrostatic simulations to extract the potential profile in the conduction channel (on top of SiO$_2$ dielectric) due to the presence of the metallic contact pads using the Finite Element Method (FEM) – a typical result for the exact geometry of device g28m6 is shown in figure~\ref{Fig:figure1}. The simulations were performed using COMSOL Multi-physics 3.5 software environment. In the simulations a potential of 10~V was applied between the contact pads and the back gate. It can be seen that the potential is quite small in the region very close to the metallic contact pads. The section of graphene monolayer lying in this region thus practically experiences no gate voltage – consequently the resistance of this section is independent of $V_g$. This physics is captured in equation~\ref{Eqn:dassharme} through the introduction of the gate independent resistance term $R_c$. Although we have included the term $R_c$ in equation~\ref{Eqn:dassharme} for the sake of generality we have noticed that setting $R_c$ to zero does not significantly affect the values of $n_0$ and mobilities (to within a few percentage) extracted from the fits. In fact, for the high mobility devices g28m6 and g30m4 the best fits were obtained when $R_c$ was identically set to zero – this is expected since the high mobility graphene acts as an equipotential surface and hence the value of the gate potential is the same everywhere on it.

\begin{figure}[!h]
\begin{center}
\includegraphics[width=0.7\textwidth]{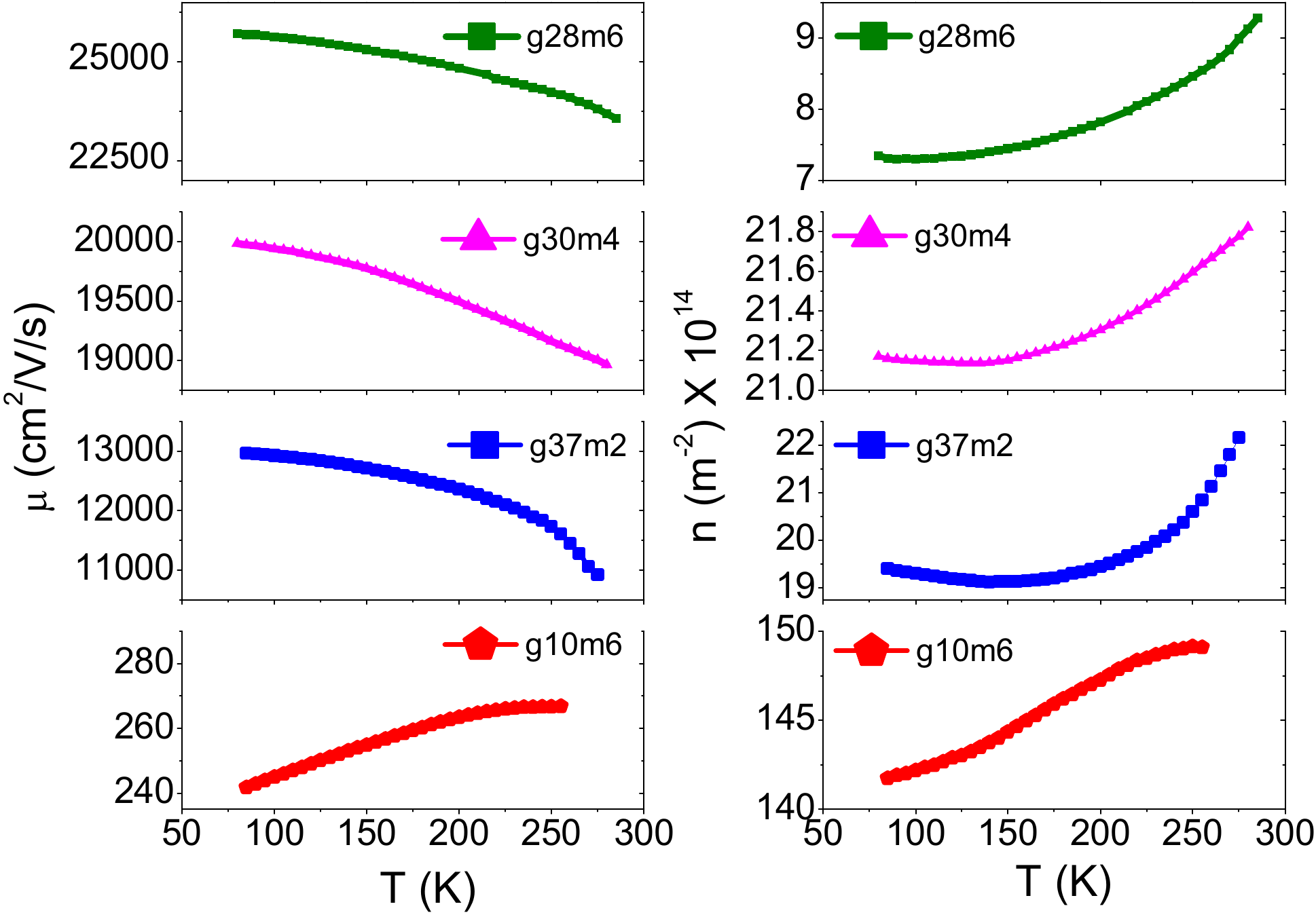}
{\caption{Plots of mobility $\mu$ \textit{vs} temperature (left panel) and of impurity number density $n_0$ \textit{vs} temperature  (right panel). Note that for all the devices $n_0$ increases with increase in temperature. For the relatively high mobility devices (g28m6, g30m4 and g37m2) $\mu$ increases with decrease in temperature  while for the lowest mobility device (g10m6) $\mu$ decreases with decrease in temperature.
\label{fig:figure3}}}
\end{center}
\end{figure}

The values of $n_0$ and $\mu$ extracted from the fits for the representative set of devices are plotted in figure~\ref{fig:figure3}. The temperature dependence of the mobility was seen to depend very strongly on the quality of the device [figure~\ref{fig:figure3}~(left panel)].  For devices having relatively large room temperature mobilities, $\mu$ was seen to increase as the temperature decreased. In contrast, for the devices having poor room temperature mobilities, $\mu$ was seen to decrease as the temperature decreased.  For all the devices studied, $n_0$ increased monotonically with temperature as is expected for the case of SLG on \sio substrate - the value of $n_0$ decreasing sharply with increase in $\mu$.

\begin{figure}[!]
\begin{center}
\includegraphics[width=0.65\textwidth]{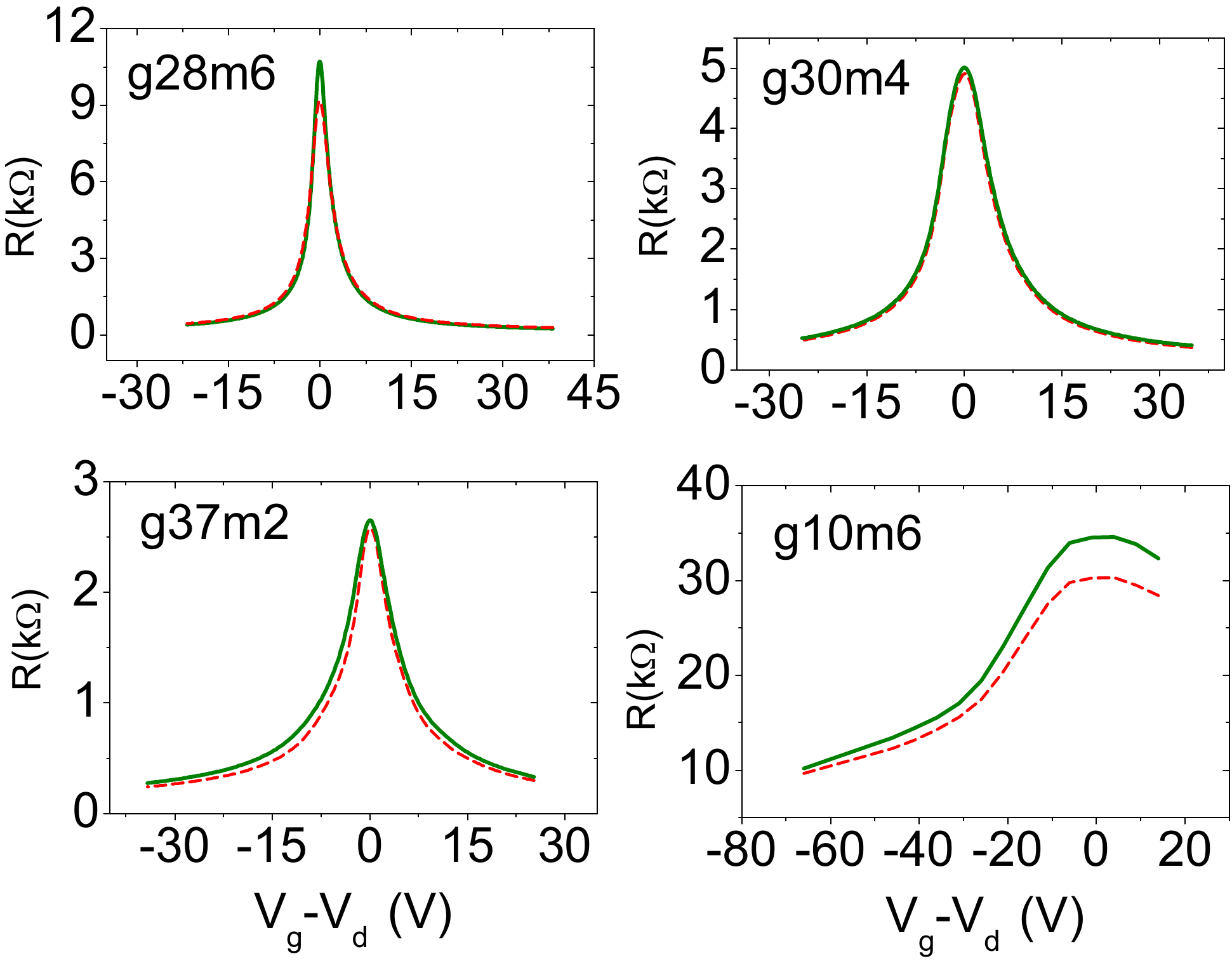}
{\caption{Plot of the resistance as a function of the reduced gate voltage (\vg-\vd) for the devices (a) g28m6, (b) g30m4, (c) g37m2 and (d) g10m6. The dotted red curve in each plot corresponds data obtained at 295~K while the solid olive curve shows the data obtained at 80~K.  \label{fig:figure4}}}
\end{center}
\end{figure}

\begin{figure*}
\begin{center}
\includegraphics[width=0.85\textwidth]{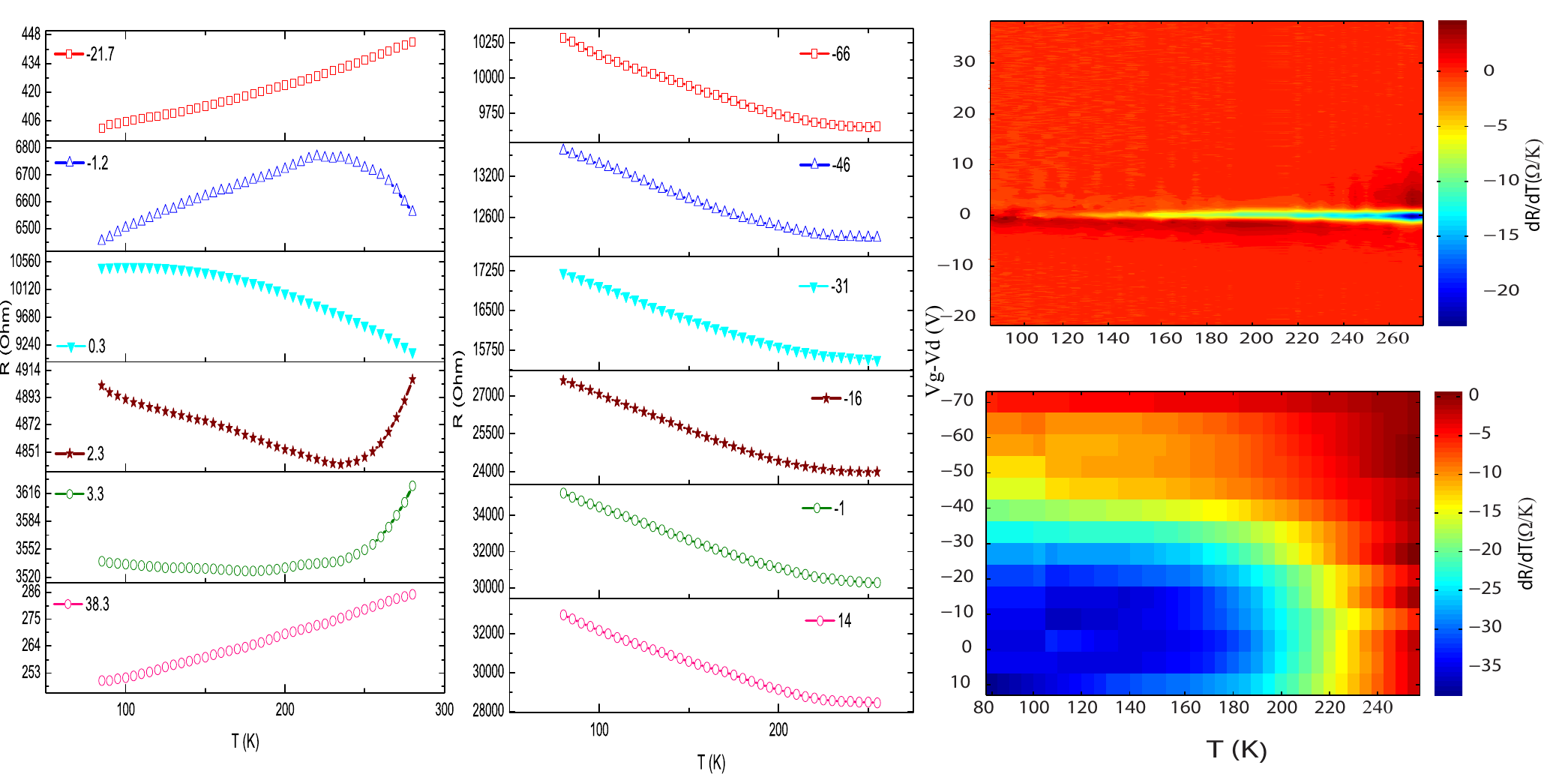}
{\caption{Resistance as a function of temperature at different gate voltages for the device g28m6 (left panel) and g10m6 (middle panel). The numbers in the plot legends refer to the value of the reduced gate voltage (\vg-\vd). (a) Plot of $dR/dT$ as a function of temperature and reducd gate voltage (\vg-\vd) for g28m6. (b) Plot of $dR/dT$ as a function of temperature and reducd gate voltage (\vg-\vd) for g10m6. \label{fig:figure5}}}
\end{center}
\end{figure*}

The resistance \textit{vs} gate voltage data for all the four  devices at a couple of representative temperatures are shown in figure~\ref{fig:figure4} (red dotted line: 295K, olive solid line: 80K). The data have been plotted as a function of $V_g-V_d$,  where \vd is the Dirac voltage identified by the maxima in the R-\vg curve. It was seen that for the high mobility devices (g28m6 and g30m4) the resistance does not change much as the temperature is swept down from 295~K to lower temperatures. In contrast, for the low mobility devices, there is a relatively large increase in the resistance as the temperature is reduced.  From the gate voltage scans we have extracted the resistance \textit{vs} temperature data at different $V_g$. Figure~\ref{fig:figure5} shows a plot of the resistance as a function of temperature for the relatively high mobility device g28m6 and for the lowest mobility device g10m6 measured at a few representative gate voltages. In the same figure we also plot the temperature coefficients of resistance $dR/dT$ measured for these two devices as a function of the  reduced gate voltage ($V_g -V_d$) and temperature. We notice from these plots that for the high mobility device g28m6 (and g30m4 - not shown here) the slope of the resistance \textit{vs} temperature curve depends on the gate voltage. When the chemical potential is away from the charge neutrality point, $dR/dT$ is positive over the entire temperature range. As the chemical potential approaches the charge neutrality point ($|$\vg-\vd$|$ small) $dR/dT$ changes sign and becomes negative. On the other hand, for device g10m6, which has a higher impurity content (as evident from a high value of $n_0$ and quite low value of $\mu$) the TCR is negative over a much larger gate voltage range. We find in all cases that the range of carrier density over which the SLG device shows an insulating behavior (dR/dT$<$0)  matches very closely with the impurity number density extracted from equation~\ref{Eqn:dassharme}. This can be understood using the following simple picture: in undoped graphene there are strong bipolar charge density fluctuations~\cite{PhysRevLett.99.176801, yacoby,PhysRevB.85.075435}. This macroscopic charge inhomogeneity leads to random  p- and n-type charge clusters~\cite{li2011disorder} in the device leading to a percolating current transport near the Dirac point. Transport in this regime is thus determined by the linkage between the percolating clusters rather than by the properties of the clusters themselves. As the chemical potential moves away from the Dirac point, eventually the number of carriers doped into the system exceeds the number density fluctuations in the system leading to a quasi-uniform chemical potential throughout the SLG device making the percolation picture irrelevant.

Before we continue to discuss the results further, we would like to discount possible quantum corrections to the measured temperature dependence of the resistance through processes like weak localization and Klein tunneling.  We have performed magnetoresistance measurements in these devices in the presence of a perpendicular magnetic field - we do not see any signatures of weak localization corrections to the resistance of any of the devices over the temperature range we are concerned with in this article (77~K-300~K). Also, it has been shown that Klein tunneling correction to the resistance of graphene monolayers becomes significant only for scattering potentials having a very steep profile~\cite{PhysRevLett.102.026807,PhysRevB.77.075420} - for example across a lithographically defined top-gate, and is not relevant for scattering across an electron-hole puddle. Hence we do not expect to have any significant contributions to the measured resistance in our devices from either of these two mechanisms.   

\section*{Discussion}

There are multiple possible sources of disorder
~\cite{tan2007measurement,chen2008charged}
scattering in an SLG device - oxide substrate, residues from polymer resists used during device fabrication, intrinsic defects of graphene, lattice distortions, ripples (due to substrate roughness) etc. Graphene has a very low density of states near the Dirac point. Consequently the screening is weak in this regime and charge impurities and trapped charges can behave as long range scatterers~\cite{macdonald.2006,PhysRevB.75.205418,PhysRevLett.97.226801,PhysRevLett.99.216602}. On the other hand, completely screened charge impurities give rise to mid-gap states which behave like short range scatterers~\cite{PhysRevB.76.205423, PhysRevB.82.081414}. The simplest phenomenological model that can be written down which incorporates the effect of both types of scatterers on transport employs the Matthiessen's rule~\cite{matthiessen} within a Boltzmann transport formalism. For concreteness, we model the effect of the completely screened charge impurities using a delta function scattering potential and the effect of the charged impurities via an unscreened Coulomb potential. Since the linear dispersion relation of SLG implies that the contribution to resistivity coming from acoustic phonon scattering is independent of carrier number density~\cite{PhysRevB.22.904,PhysRevB.76.205423}, we ignore phonon scattering in our calculations.  The Boltzmann relaxation time $\tau_{k}$ due to a scattering potential that results in an $S$-matrix $S(k,k')$ is given by~\cite{castro.2009,dassarma.2008}

\begin{eqnarray}
\frac{1}{\tau_k}=n_{i}\int{d\theta}\int{\frac{k'dk'}{(2\pi)^{2}} S(k,k')(1-\cos\theta)},
\label{Eqn:Hamiltonian}
\end{eqnarray}
\noindent
where $n_{i}$ is impurity concentration per unit area and $k$ is the magnitude of the initial momentum of the scattered particle. The integration is over all possible 
final momentum states of magnitude $k'$ making an angle $\theta$ with respect to the initial momentum.  Further,
the transition rate $S(k,k')$ is given, in the Born approximation,
by
\begin{eqnarray}
 S(k,k')=2\pi|H_{k,k'}|^{2}\frac{1}{v_{F}}\delta(k'-k),
\end{eqnarray}
where $v_{F}$ is the fermi velocity  and  $H_{k,k'}$, the transition matrix elements between states $|k'>$ and $|k>$ is defined as
\begin{eqnarray}
 H_{k,k'}=\int{dr \psi^{\ast}_{k'}(r)U_{s}(r)\psi_{k}(r)},
\end{eqnarray}
where $U_{s}(r)$ is the scattering potential and $\psi_{k}(r)$ is the electronic spinor wave function of a clean graphene sheet. For a  short range scattering potential with $U_{s}(r)=v_{0}\delta(r)$,~\cite{castro.2009} where $v_{0}$ is the strength of delta potential,
\begin{eqnarray}
 \tau_{k}=\frac{4v_{F}}{n_{i}v_{0}^{2}k}.
\end{eqnarray}
On the other hand, for the Coulomb potential with $U_{s}(r)=\frac{eQ}{4\pi\epsilon_{0}\epsilon _rr}$~\cite{castro.2009}
\begin{eqnarray}
 \tau_{k}=\frac{v_{F}k}{u_{0}^{2}}
\end{eqnarray}
where, $u_{0}^{2}=n_{i}Q^{2}e^{2}/(16\epsilon_{0}^{2}\epsilon _r^{2}r^{2})$, $Q$ is the impurity charge and $\epsilon_r$ is the  relative electric permittivity.\\

Graphene has a hexagonal Brillouin zone. At its six corners, the energy dispersion is given by $E(k)=\pm\hbar v_{F}|k|$ and the density of states is $g(E)=\frac{2|E|}{\pi\hbar^{2}v_{F}^{2}}$
\cite{patel2012thermoelectricity}. In Boltzmann transport theory, the conductivity($\sigma$) is given by~\cite{dassarma.2009},
\begin{eqnarray}
 \sigma(T)=\frac{e^{2}v_{F}^{2}}{2}\int{d\epsilon g(\epsilon)\tau(\epsilon)(-\frac{\partial f}{\partial \epsilon})}
\end{eqnarray}
where $f(\epsilon_{k})=\frac{1}{1+\exp[(\epsilon_{k}-\epsilon_{F})/K_{B}T]}$ is the Fermi-Dirac distribution,  $\epsilon_{F}$ is chemical potential and $T$ is the temperature. The linear dispersion relation of graphene implies that $\tau(E)\propto|E|$ for unscreened coulomb potential~\cite{patel2012thermoelectricity} and  $\tau(E)\propto\frac{1}{|E|}$ for completely screened delta potential.
It can be shown that the unscreened Coulomb potential alone results in insulating transport characterized by $dR/dT<0$. On the other hand, the delta function potential by itself results in metallic transport with $dR/dT>0$. It is thus natural to attempt to describe the experimental results using a combination of the two mechanisms, the physical motivation for which was mentioned earlier. We thus define an effective scattering time $\tau$, which is a combination of $\tau_{couloumb}$ and $\tau_{delta}$ by means of Mathhiessen's rule as,
\begin{eqnarray}
\frac{1}{\tau}= \frac{w_1}{\tau_{couloumb}}+\frac{w_2}{\tau_{delta}}.
\end{eqnarray}
$w_1$ and $w_2$, the weights of the two potentials are determined by fits to the experimental data. Specifically, we fix $w_1$ and $w_2$ for a sample at a particular temperature by fitting to the resistance as a function of the chemical potential $\epsilon_{F}$.\\

\begin{figure}
\begin{center}
\includegraphics[width=0.85\textwidth]{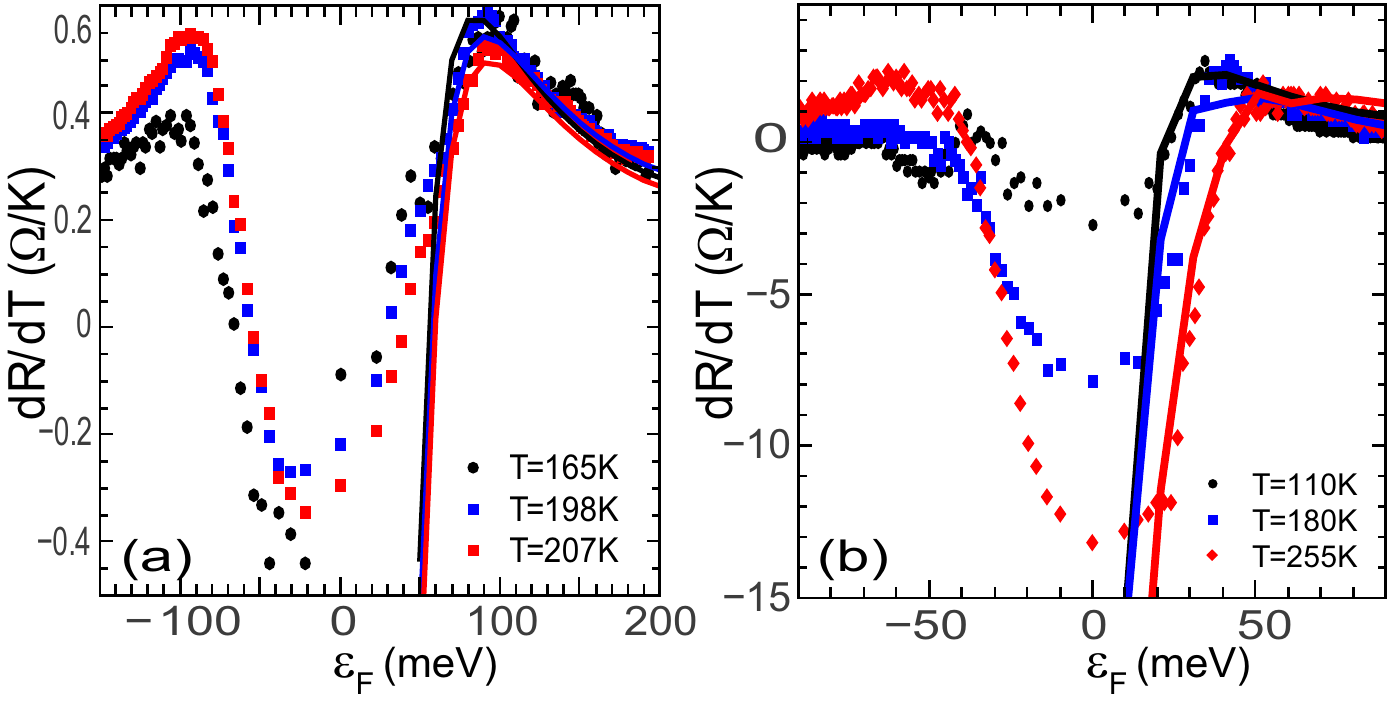}
{\caption{Variation of $dR/dT$ with the chemical potential $\epsilon_F$ at different temperatures. The scatter points are the measured experimental data while the solid lines are best fits from our theoretical calculation. The data are presented for two different devices - (a) relatively low mobility device g7m5 and (b) relatively high mobility device g28m6. \label{Fig:figure6}}}
\end{center}
\end{figure}

In figure~\ref{Fig:figure6} we have plotted the $dR/dT$ as a function of the chemical potential $\epsilon_F$ for two different devices - (a) g7m5 which is a relatively low mobility device and (b) g28m6 which is a high mobility device. It can be seen that the fits to the experimental data using our semi-classical theory near the Dirac point works much better in the case of the high mobility device than for the device with low mobility. This can be understood as follows: the semi-classical Boltzmann transport calculations presented here are based on the one-band approximation
~\cite{hwangprl.2007, macdonald.2006}.
It is generally expected that such semi-classical models should break down in the vicinity of zero doping as $\epsilon_F \tau$ becomes comparable to $\hbar$. Having said that, it was recently shown that for short-range scatterers (which, as shown below, are the primary sources of scattering for high mobility SLG) the carrier momentum relaxation time diverges near the zero doping limit~\cite{PhysRevLett.98.076602}. Thus the value of $\epsilon_F \tau$ (and consequently the validity of semi-classical treatment) depends not only on the carrier concentration but also critically on the scattering mechanism~\cite{PhysRevLett.99.216602}. 

$w_1$ and $w_2$ are temperature and sample dependent parameters and can indeed result in a non-monotonic behaviour of the resistivity - going from metallic to insulating as a function of increasing temperature as seen in figures~\ref{Fig:figure6}(a) and (b).  As can be seen from figure~\ref{Fig:figure6}, this simple theory can also explain the change in sign of $dR/dT$ with changing chemical potential. We emphasize that we use this model purely phenomenologically to describe the experimental data and do not attempt to explain why the coefficients $w_1$ and $w_2$ should have the observed temperature dependences, deferring it to a future work.

\begin{figure}
\begin{center}
\includegraphics[width=0.75\textwidth]{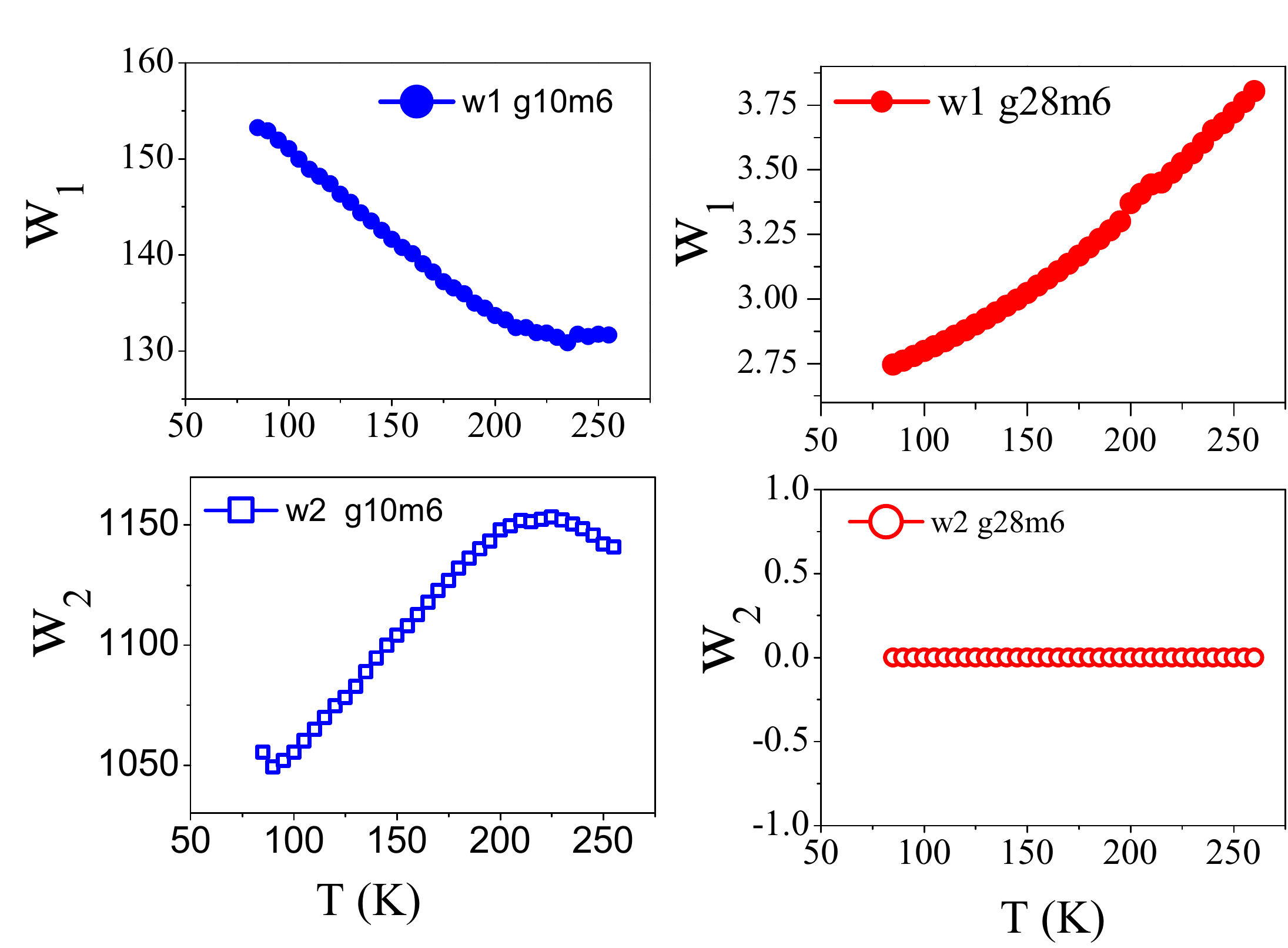} 
{\caption{Variation of $w_{1}$ (weight factor associated with Coulomb potential) and $w_{2}$  (weight factor associated with delta potential) with temperature for devices g10m6 and g28m6.\label{Fig:figure7}}}
\end{center}
\end{figure}

From the fits to the experimentally measured $dR/dT$ \textit{vs} $\epsilon_F$ data, we find that for the relatively high mobility devices g28m6 and g30m4, $w_1$ (which is in some sense a measure of impurity concentrations) increases with temperature as shown in figure~\ref{Fig:figure6}. From figure~\ref{fig:figure3} it can be seen that for both these devices $n_0$ increases with an increase in temperature. In general we expect that as impurity concentration increases the mobility should decrease. This is exactly what we observe in the experiments for these two and for all other high mobility devices that we have studied. Except for gate voltages very close to the Dirac point, these devices show a metallic behaviour with $dR/dT>0$ at all temperatures. On the other hand the low mobility devices like g10m6 exhibit $dR/dT<0$ over extended ranges of temperature at all values of gate voltage. For these samples, we observe that both $n_0$ and $\mu$ increase with increasing temperature. For such low mobility  devices, $w_2>>w_1$ and hence the variation of  $n_0$  with temperature is controlled by the weight factor $w_2$ (see figure~\ref{Fig:figure7}). From our fits we find that $w_2$ increases with an increase in temperature leading to an increase of $n_0$ with increasing temperature.  The negative values of TCR for these devices over extended ranges of temperature indicates that the relevant scattering mechanism governing the mobility of the two devices is ionized impurities (which leads to insulating behaviour). As a result, we expect that the behaviour of mobility in these devices will be mainly governed by  $w_1$. We find from our fits that $w_1$ decreases with increasing temperature causing the mobility (which goes inversely as $w_1$) to increase as the temperature increases. 

In principle, the resistivity data can also be explained with either one of the two scattering mechanisms with an appropriate temperature dependent weight in front. This weight 
would then correspond to the concentration of impurities causing the scattering. However, the simultaneous increase of impurity concentration 
and mobility with $T$ (as seen in the low mobility samples) cannot be explained this way and two different types of scattering mechanisms become essential in our 
phenomenological model. 

\section*{Conclusion}
In summary, we have investigated the temperature dependence of the resistance of single layer graphene devices on SiO$_2$ substrates. The temperature dependence of the electrical transport of these SLG devices could be explained within a Boltzmann transport formalism involving two independent scattering mechanisms - (i) long range Coulomb scattering and (ii) short range delta potential. In all our high mobility devices an insulating behavior was seen only very close to the Dirac point in the doping region determined by bipolar charge density fluctuations.  We also find that for high mobility devices the strength of long range scattering potential is much smaller than that seen for the low mobility devices. This observation corroborates with the finding that graphene deposited on substrates with strong screening properties like the ferroelectric oxide Pb(Zr$_{0.2}$Ti$_{0.8}$)O$_3$ (PZT) led to a large increase in carrier mobilities~\cite{PhysRevLett.102.136808}.  On the other hand, for the low mobility devices electrical transport properties are determined by both types of scattering potentials - long range due to charged ionized impurities and short range due to completely screened charge impurities. We find that long range Coulomb scatterers can lead to low carrier mobility and insulating behaviour seen even in the high doping region in these devices. This is in contrast to the metallic conduction and high carrier mobilities seen in devices where coulomb interactions were small.

\bibliography{graphene_dRdT}

\section*{Methods}

Graphene flakes were mechanically exfoliated on 300 nm SiO$_2$ substrates from natural graphite crystals. Single layer flakes are identified by the colour contrast on optical microscope and by Raman spectroscopy. In some cases single layer graphene was also confirmed by the position of quantum hall plateaus. Electrical contact on the identified monolayer graphene flakes were defined by standard electron beam lithography technique followed by thermal deposition of Cr (5nm)/Au(70nm) metal contacts. In figure~\ref{Fig:figure1} we show a schematic of 4 probe graphene device on Si/SiO$_2$ substrate.

\section*{Acknowledgments}

We acknowledge funding from Nanomission, Department of Science \& Technology (DST), Govt. of India and HRDG, CSIR, Govt. of India. We thank Ramya Nagarajan for help with the simulations.

\section*{Author contributions statement}

A.B. conceived the experiments,  S.S., K.R.A. and A.S. conducted the experiments, S.S., K,R.A., R.M., S.M. and A.B. analyzed the results, R.M. and S.M. performed the calculations.  All authors reviewed the manuscript. 

\section*{Competing financial interests}

The authors declare no competing financial interests. 

\end{document}